\documentclass{article}
\def\Ref#1{(\ref{#1})}
\usepackage{amsmath}
\usepackage{amssymb}
\usepackage{cite}
\begin{document}
\begin{titlepage}
\noindent{\large \bfseries Exactly solvable models through the
generalized empty interval method: multi-species and
more-than-two-site interactions} \vskip 2 cm
\begin{center}{Amir~Aghamohammadi{\footnote
{mohamadi@azzahra.ac.ir}} \& Mohammad~Khorrami{\footnote
{mamwad@mailaps.org}}} \vskip 5 mm

{\it{Department of Physics, Alzahra~University,
             Tehran 19938-91167, Iran. }}
\end{center}

\begin{abstract}
\noindent Multi-species reaction-diffusion systems, with
more-than-two-site interaction on a one-dimensional lattice are
considered. Necessary and sufficient constraints on the
interaction rates are obtained, that guarantee the closedness of
the time evolution equation for $E^{\mathbf a}_n(t)$'s, the
expectation value of the product of certain linear combination of
the number operators on $n$ consecutive sites at time $t$.
\end{abstract}
\begin{center} {\bf PACS numbers:} 05.40.-a, 02.50.Ga

{\bf Keywords:} reaction-diffusion, generalized empty-interval
method, next-nearest-neighbor interaction, multi-species
\end{center}

\end{titlepage}
\newpage
\section{Introduction}
The study of the reaction-diffusion systems, has been an
attractive area. A reaction-diffusion system consists of a
collection of particles (of one or several species) moving and
interacting with each other with specific probabilities (or rates
in the case of continuous time variable). In the so called
exclusion processes, any site of the lattice the particles move
on, is either vacant or occupied by one particle. The aim of
studying such systems, is of course to calculate the time
evolution of such systems. But to find the complete time evolution
of a reaction-diffusion system, is generally a very difficult (if
not impossible) task.

Reaction-diffusion systems have been studied using various methods:
analytical techniques, approximation methods, and simulation. The
success of the approximation methods, may be different in
different dimensions, as for example the mean field techniques,
working good for high dimensions, generally do not give correct
results for low dimensional systems. A large fraction of
analytical studies, belong to low-dimensional (specially
one-dimensional) systems, as solving low-dimensional systems
should in principle be easier
\cite{ScR,ADHR,KPWH,HS1,PCG,HOS1,HOS2,AL,AKK,RK,RK2,AKK2,AAMS,AM1}.

Various classes of reaction-diffusion systems are called
exactly-solvable, in different senses. In \cite{AA,RK3,RK4},
integrability means that the $N$-particle conditional
probabilities' S-matrix is factorized into a product of 2-particle
S-matrices. This is related to the fact that for systems solvable
in this sense, there are a large number of conserved quantities.
In \cite{BDb,BDb1,BDb2,BDb3,Mb,HH,AKA,KAA,MB,AAK}, solvability
means closedness of the evolution equation of the empty intervals
(or their generalization).

The empty interval method (EIM) has been used to analyze the one
dimensional dynamics of diffusion-limited coalescence
\cite{BDb,BDb1,BDb2,BDb3}. Using this method, the probability that
$n$ consecutive sites are empty has been calculated.  For the
cases of finite reaction-rates, some approximate solutions have
been obtained. EIM has been also generalized to study the kinetics
of the $q$-state one-dimensional Potts model in the
zero-temperature limit \cite{Mb}.

In \cite{AKA},  all the one dimensional reaction-diffusion models
with nearest neighbor interactions which can be exactly solved by
EIM have been studied. EIM has also been used to study a  model
with next nearest neighbor interaction \cite{HH}. In \cite{KAA},
exactly solvable models through the empty-interval method, for
more-than-two-site interactions were studied. There, conditions
were obtained which are sufficient for  a one-species system to be
solvable through the EIM.  In \cite{MB}, the conventional EIM has
been extended to a more generalized form. Using this extended
version, a model has been studied, which can not be solved by
conventional EIM.

In a recent article \cite{AAK},  we considered  nearest-neighbor
multi-species models on a one-dimensional lattice. we obtained
necessary and sufficient conditions on the reaction rates, so that
the time evolution equation for $E^{\mathbf a}_{k,n}(t)$ is
closed. Here $E^{\mathbf a}_{k,n}(t)$ is the expectation of the
product of a specific linear combination of the number operators
(corresponding to different species) at $n$ consecutive sites
beginning from the $k$-th site. All single-species left-right
symmetric reaction-diffusion systems solvable through the
generalized empty-interval method (GEIM), were classified. In this
article, the method introduced in \cite{AAK} is generalized to
more than two-site interactions.

In section 2, multi-species systems with three-site interactions
are investigated, which are solvable through the GEIM. Conditions
necessary and sufficient for closedness of the evolution equation
of $E^{\mathbf a}_{k,n}(t)$ are obtained, and the evolution
equation is explicitly given.

In section 3, the more general case of multi-species systems with
$k$-site interactions, solvable through the GEIM, is investigated.
Again, conditions necessary and sufficient for
closedness of the evolution equation of $E^{\mathbf a}_{k,n}(t)$
as well as the evolution equation itself, are obtained.

Finally, in section 4, as an example a specific model with a
three-site interaction is introduced and exactly solved.

\section{Three-site interactions}
Consider a periodic lattice with $L+1$ sites. Each site is either
empty, or occupied with a particle of one of $p$ possible species.
Denote by $N_i^\alpha$, the number operator of the particles of
type $\alpha$ at the site $i$. $\alpha=p+1$ is regarded as a
vacancy. $N^\alpha_i$ is equal to one, if the site $i$ is occupied
by a particle of type $\alpha$. Otherwise, $N^\alpha_i$ is zero.
We also have a constraint
\begin{equation}\label{1}
s_\alpha N^\alpha_i=1,
\end{equation}
where ${\mathbf s}$ is a covector the components of which
($s_\alpha$'s) are all equal to one. The constraint \Ref{1},
simply says that every site, either is occupied by a particle of
one type, or is empty. A representation for these observables is
\begin{equation}\label{2}
N_i^\alpha:=\underbrace{1\otimes\cdots\otimes 1}_{i-1}\otimes
N^\alpha\otimes\underbrace{1\otimes\cdots\otimes 1}_{L+1-i},
\end{equation}
where $N^\alpha$ is a diagonal $(p+1)\times(p+1)$ matrix the only
nonzero element of which is the $\alpha$'th diagonal element, and
the operators 1 in the above expression are also
$(p+1)\times(p+1)$ matrices. It is seen that the constraint
\Ref{1} can be written as
\begin{equation}\label{3}
{\mathbf s}\cdot{\mathbf N}=1,
\end{equation}
where ${\mathbf N}$ is a vector the components of which are
$N^\alpha$'s. The state of the system is characterized by a vector
\begin{equation}\label{4}
{\mathbf P}\in\underbrace{{\mathbb V}\otimes\cdots\otimes{\mathbb
V}}_{L+1},
\end{equation}
where ${\mathbb V}$ is a $(p+1)$-dimensional vector space. All the
elements of the vector ${\mathbf P}$ are nonnegative, and
\begin{equation}\label{5}
{\mathbf S}\cdot{\mathbf P}=1.
\end{equation}
Here ${\mathbf S}$ is the tensor-product of $L+1$ covectors
${\mathbf s}$.

As the number operators $N^\alpha_i$ are zero or one (and hence
idempotent), the most general observable of such a system is the
product of some of these number operators, or a sum of such terms.

The evolution of the state of the system is given by
\begin{equation}\label{6}
\dot{\mathbf P}={\mathcal H}\;{\mathbf P},
\end{equation}
where the Hamiltonian ${\mathcal H}$ is stochastic, by which it is
meant that its non-diagonal elements are nonnegative and
\begin{equation}\label{7}
{\mathbf S}\; {\mathcal H}=0.
\end{equation}
The interaction is a next-nearest-neighbor interaction:
\begin{equation}\label{8}
{\mathcal H}=\sum_{i=1}^{L+1}H_{i,i+1,i+2},
\end{equation}
where
\begin{equation}\label{9}
H_{i,i+1,i+2}:=\underbrace{1\otimes\cdots\otimes 1}_{i-1}\otimes H
\otimes\underbrace{1\otimes\cdots\otimes 1}_{L-i-1}.
\end{equation}
(It has been assumed that the sites of the system are identical,
that is, the system is translation-invariant. Otherwise $H$ in the
right-hand side of \Ref{9} would depend on $i$.) The three-site
Hamiltonian $H$ is stochastic, that is, its non-diagonal elements
are nonnegative, and the sum of the elements of each of its
columns vanishes:
\begin{equation}\label{10}
({\mathbf s}\otimes{\mathbf s}\otimes{\mathbf s})H=0.
\end{equation}

Now consider a certain class of observables, namely
\begin{equation}\label{11}
{\mathcal E}^{\mathbf a}_{k,n}:=\prod_{l=k}^{k+n-1}({\mathbf
a}\cdot{\mathbf N}_l),
\end{equation}
where ${\mathbf a}$ is a specific $(p+1)$-dimensional covector,
and ${\mathbf N}_i$ is a vector the components of which are the
operators $N_i^\alpha$. We want to find criteria for $H$, so that
the evolutions of the expectations of ${\mathcal E}^{\mathbf
a}_{k,n}$'s are closed, that is, the time-derivative of their
expectation is expressible in terms of the expectations of
${\mathcal E}^{\mathbf a}_{k,n}$'s themselves. Denoting the
expectations of these observables by $E^{\mathbf a}_{k,n}$,
\begin{equation}\label{12}
E^{\mathbf a}_{k,n}:={\mathbf S}\;{\mathcal E}^{\mathbf
a}_{k,n}{\mathbf P}.
\end{equation}
For $1<n<L$, we have
\begin{align}\label{13}
\dot E^{\mathbf a}_{k,n}=&{\mathbf S}\;{\mathcal E}^{\mathbf
a}_{k,n}
{\mathcal H}\;{\mathbf P},\nonumber\\
=&\sum_{l=-1}^{n}{\mathbf S}\;{\mathcal E}^{\mathbf
a}_{k,n}H_{k+l-1,k+l,k+l+1}
\;{\mathbf P}\nonumber\\
=&{\mathbf S}\;{\mathcal E}^{\mathbf a}_{k,n}H_{k-2,k-1,k}\;{\mathbf P}\nonumber\\
&+{\mathbf S}\;{\mathcal E}^{\mathbf a}_{k,n}H_{k-1,k,k+1}\;{\mathbf P}\nonumber\\
&+\sum_{l=1}^{n-2}{\mathbf S}\;{\mathcal E}^{\mathbf
a}_{k,n}H_{k-1+l,k+l,k+l+1}
\;{\mathbf P}\nonumber\\
&+{\mathbf S}\;{\mathcal E}^{\mathbf a}_{k,n}H_{k+n-2,k+n-1,k+n}\;{\mathbf P}\nonumber\\
&+{\mathbf S}\;{\mathcal E}^{\mathbf a}_{k,n}H_{k+n-1,k+n,k+n+1}\;{\mathbf P}.
\end{align}
In the right-hand side of the last equality, the first two and the
last two terms are contributions of the interactions at the
boundaries of the block, while the summation term is the
contribution of the sites in the bulk.

To proceed, one may use the following identity
\begin{equation}\label{14}
{\mathbf s}({\mathbf b}\cdot{\mathbf N})={\mathbf b},
\end{equation}
using which, one arrives at
\begin{equation}\label{15}
({\mathbf s}\otimes{\mathbf s }\otimes{\mathbf s})[({\mathbf
a}\cdot{\mathbf N})\otimes ({\mathbf a}\cdot{\mathbf N})\otimes
({\mathbf a}\cdot{\mathbf N})H]=({\mathbf a}\otimes{\mathbf
a}\otimes{\mathbf a})H.
\end{equation}
Demanding that the bulk terms in (\ref{13}) be expressible in
terms of the expectations of ${\mathcal E}^{\mathbf a}_{k,n}$'s
themselves, one arrives at the following condition for the bulk
terms.

\begin{equation}\label{17}
({\mathbf a}\otimes{\mathbf a}\otimes{\mathbf a})H=\lambda
({\mathbf a}\otimes{\mathbf a}\otimes{\mathbf a}),
\end{equation}
for some constant $\lambda$. By similar arguments, the condition
coming from the left boundary terms of  (\ref{13}) can be
obtained:
\begin{align}\label{18}
({\mathbf s}\otimes{\mathbf s}\otimes{\mathbf a}\otimes{\mathbf a})[(1\otimes H)
+(H\otimes 1)]=
&\nonumber \\
\mu_{\mathrm L}{\mathbf s}\otimes{\mathbf s}\otimes{\mathbf s}\otimes{\mathbf s}&+
\nu_{\mathrm L}{\mathbf s}\otimes{\mathbf s}\otimes{\mathbf s}\otimes{\mathbf a}+
\gamma_{\mathrm L}{\mathbf s}\otimes{\mathbf s}\otimes{\mathbf a}\otimes{\mathbf a}
\nonumber \\
&+ \delta_{\mathrm L}{\mathbf s}\otimes{\mathbf a}\otimes{\mathbf a}\otimes{\mathbf a}+
\phi_{\mathrm L}{\mathbf a}\otimes{\mathbf a}\otimes{\mathbf a}\otimes{\mathbf a},
\end{align}
and finally, the condition coming from the right boundary terms of
(\ref{13}) is
\begin{align}\label{19}
({\mathbf a}\otimes{\mathbf a}\otimes{\mathbf s}\otimes{\mathbf s})[(1\otimes H)
+(H\otimes 1)]=
&\nonumber \\
\mu_{\mathrm R}{\mathbf s}\otimes{\mathbf s}\otimes{\mathbf s}\otimes{\mathbf s}&+
\nu_{\mathrm R}{\mathbf a}\otimes{\mathbf s}\otimes{\mathbf s}\otimes{\mathbf s}+
\gamma_{\mathrm R}{\mathbf a}\otimes{\mathbf a}\otimes{\mathbf s}\otimes{\mathbf s}
\nonumber \\
&+ \delta_{\mathrm R}{\mathbf a}\otimes{\mathbf a}\otimes{\mathbf
a}\otimes{\mathbf s}+ \phi_{\mathrm R}{\mathbf a}\otimes{\mathbf
a}\otimes{\mathbf a}\otimes{\mathbf a},
\end{align}
where the multipliers in the right-hand sides of (\ref{18}) and
(\ref{19}) are constants. Arranging all these together, the
evolution equation for $E^{\mathbf a}_{k,n}$, for $1<n<L$, becomes
\begin{align}\label{19-19}
\dot E^{\mathbf a}_{k,n}=&\mu_{\mathrm L}E^{\mathbf a}_{k+2,n-2}+
\mu_{\mathrm R}E^{\mathbf a}_{k,n-2}+\nu_{\mathrm L}E^{\mathbf
a}_{k+1,n-1}+\nu_{\mathrm R}E^{\mathbf a}_{k,n-1}\nonumber\\
&+[\gamma_{\mathrm L}+(n-2)\lambda+\gamma_{\mathrm R}]E^{\mathbf
a}_{k,n}\nonumber\\
&+\delta_{\mathrm L}E^{\mathbf a}_{k-1,n+1}+\delta_{\mathrm
R}E^{\mathbf a}_{k,n+1}+\phi_{\mathrm L}E^{\mathbf
a}_{k-2,n+2}+\phi_{\mathrm R}E^{\mathbf a}_{k,n+2}.
\end{align}
In general, even if the initial conditions are not translationally
invariant, one can solve the above equation. However, assuming
that the initial conditions are translationally invariant,
simplifies the calculations. Assuming that the initial conditions
are translationally invariant, as the dynamics is also
translationally invariant, $E^{\mathbf a}_{k,n}$ is independent of
$k$, and so one arrives at
\begin{align}\label{19-2}
\dot E^{\mathbf a}_{n}=&(\mu_{\mathrm L}+ \mu_{\mathrm
R})E^{\mathbf a}_{n-2}+(\nu_{\mathrm L}+\nu_{\mathrm R})E^{\mathbf
a}_{n-1}+[\gamma_{\mathrm L}+(n-2)\lambda+\gamma_{\mathrm R}]
E^{\mathbf a}_{n}\nonumber\\
&+(\delta_{\mathrm L}+\delta_{\mathrm
R})E^{\mathbf a}_{n+1}+(\phi_{\mathrm L}+\phi_{\mathrm
R})E^{\mathbf a}_{n+2}.
\end{align}

The cases of zero-, one-, $L$-, and $(L+1)$-point functions should
be considered separately. The zero-point function is equal to one.
In the case of the one-point function, one arrives at an
additional condition
\begin{align}\label{18-2}
({\mathbf s}\otimes{\mathbf s}\otimes{\mathbf a}\otimes{\mathbf
s}\otimes{\mathbf s})[(1\otimes 1\otimes H) +(1\otimes H\otimes
1)+( H\otimes 1\otimes 1)]=
\nonumber \\
\mu_{0}({\mathbf s}\otimes{\mathbf s}\otimes{\mathbf
s}\otimes{\mathbf s}\otimes{\mathbf s})+ \mu_{4}({\mathbf
s}\otimes{\mathbf s}\otimes{\mathbf a}\otimes{\mathbf
s}\otimes{\mathbf s})\nonumber
\\ + \mu_{8}({\mathbf s}\otimes{\mathbf a}\otimes{\mathbf
s}\otimes{\mathbf s}\otimes{\mathbf s})+ \mu_{16}({\mathbf
a}\otimes{\mathbf s}\otimes{\mathbf s}\otimes{\mathbf
s}\otimes{\mathbf s})&\nonumber \\ + \mu_{12}({\mathbf
s}\otimes{\mathbf a}\otimes{\mathbf a}\otimes{\mathbf
s}\otimes{\mathbf s})+ \mu_{24}({\mathbf a}\otimes{\mathbf
a}\otimes{\mathbf s}\otimes{\mathbf s}\otimes{\mathbf s})
&\nonumber \\ + \mu_{28}({\mathbf a}\otimes{\mathbf
a}\otimes{\mathbf a}\otimes{\mathbf s}\otimes{\mathbf s}) +
\mu_{2}({\mathbf s}\otimes{\mathbf s}\otimes{\mathbf
s}\otimes{\mathbf a}\otimes{\mathbf s}) &\nonumber \\ +
\mu_{6}({\mathbf s}\otimes{\mathbf s}\otimes{\mathbf
a}\otimes{\mathbf a}\otimes{\mathbf s}) + \mu_{14}({\mathbf
s}\otimes{\mathbf a}\otimes{\mathbf a}\otimes{\mathbf
a}\otimes{\mathbf s}) &\nonumber \\ + \mu_{7}({\mathbf
s}\otimes{\mathbf s}\otimes{\mathbf a}\otimes{\mathbf
a}\otimes{\mathbf a}) + \mu_{3}({\mathbf s}\otimes{\mathbf
s}\otimes{\mathbf s}\otimes{\mathbf a}\otimes{\mathbf a})
&\nonumber \\ + \mu_{1}({\mathbf s}\otimes{\mathbf
s}\otimes{\mathbf s}\otimes{\mathbf s}\otimes{\mathbf a}).
\end{align}
Then the evolution equation for $E^{\mathbf a}_{1}$ is
\begin{align}
\dot E^{\mathbf a}_{1}=&(\mu_{7}+\mu_{14}+\mu_{28})E^{\mathbf
a}_{3} +(\mu_{3}+\mu_{6}+\mu_{12}+\mu_{24})E^{\mathbf
a}_{2}\nonumber
\\ &+(\mu_{1}+\mu_{2}+\mu_{4}+\mu_{8}+\mu_{16})E^{\mathbf a}_{1}+\mu_{0}.
\end{align}

In the case $n=L$, one arrives at the following additional
condition
\begin{align}\label{18-3}
({\mathbf a}\otimes{\mathbf a}\otimes{\mathbf s}\otimes{\mathbf
a}\otimes{\mathbf a})[(1\otimes 1\otimes H) +(1\otimes H\otimes
1)+( H\otimes 1\otimes 1)]=
\nonumber \\
\nu_{31}({\mathbf a}\otimes{\mathbf a}\otimes{\mathbf
a}\otimes{\mathbf a}\otimes{\mathbf a})+ \nu_{27}({\mathbf
a}\otimes{\mathbf a}\otimes{\mathbf s}\otimes{\mathbf
a}\otimes{\mathbf a})\nonumber
\\ + \nu_{23}({\mathbf a}\otimes{\mathbf s}\otimes{\mathbf
a}\otimes{\mathbf a}\otimes{\mathbf a})+ \nu_{15}({\mathbf
s}\otimes{\mathbf a}\otimes{\mathbf a}\otimes{\mathbf
a}\otimes{\mathbf a})&\nonumber \\ + \nu_{19}({\mathbf
a}\otimes{\mathbf s}\otimes{\mathbf s}\otimes{\mathbf
a}\otimes{\mathbf a})+ \nu_{7}({\mathbf s}\otimes{\mathbf
s}\otimes{\mathbf a}\otimes{\mathbf a}\otimes{\mathbf a})
&\nonumber \\ + \nu_{3}({\mathbf s}\otimes{\mathbf
s}\otimes{\mathbf s}\otimes{\mathbf a}\otimes{\mathbf a}) +
\nu_{29}({\mathbf a}\otimes{\mathbf a}\otimes{\mathbf
a}\otimes{\mathbf s}\otimes{\mathbf a}) &\nonumber \\ +
\nu_{25}({\mathbf a}\otimes{\mathbf a}\otimes{\mathbf
s}\otimes{\mathbf s}\otimes{\mathbf a}) + \nu_{17}({\mathbf
a}\otimes{\mathbf s}\otimes{\mathbf s}\otimes{\mathbf
s}\otimes{\mathbf a}) &\nonumber \\ + \nu_{24}({\mathbf
a}\otimes{\mathbf a}\otimes{\mathbf s}\otimes{\mathbf
s}\otimes{\mathbf s}) + \nu_{28}({\mathbf a}\otimes{\mathbf
a}\otimes{\mathbf a}\otimes{\mathbf s}\otimes{\mathbf s})
&\nonumber \\ + \nu_{30}({\mathbf a}\otimes{\mathbf
a}\otimes{\mathbf a}\otimes{\mathbf a}\otimes{\mathbf s}).
\end{align}
Then the evolution equation for $E^{\mathbf a}_{L}$ is
\begin{align}
\dot E^{\mathbf a}_{L}=&(\nu_{3}+\nu_{17}+\nu_{24})E^{\mathbf
a}_{L-2}+(\nu_{7}+\nu_{19}+\nu_{25}+\nu_{28})E^{\mathbf a}_{L-1}
\nonumber \\
&+(\nu_{15}+\nu_{23}+\nu_{27}+\nu_{29}+\nu_{30})E^{\mathbf
a}_{L}+\nu_{31}E^{\mathbf a}_{L+1}.
\end{align}

The evolution equation for the case $n=L+1$ is simply
\begin{equation}\label{19-3}
\dot E^{\mathbf a}_{L+1}=\lambda (L+1)E^{\mathbf a}_{L+1}.
\end{equation}

\section{$k$-site interactions}
Now, let's go further and consider the more general case of
$k$-site interactions. The problem of single-species $k$-site
interactions solvable in the framework of empty interval method
was addressed in \cite{KAA}. In that article necessary conditions
for a system to be solvable through the EIM was obtained. Here we
obtain the necessary and sufficient conditions for a system to be
solvable through the GEIM. To fix notation, let's introduce
\begin{align}\label{26}
    {\cal A}^{ijl}:&={\mathbf s}^{\otimes i}\otimes
    {\mathbf a}^{\otimes j}\otimes {\mathbf s}^{\otimes l},\cr
    {\cal B}^{ijl}:&={\mathbf a}^{\otimes i}\otimes
    {\mathbf s}^{\otimes j}\otimes {\mathbf a}^{\otimes l},
\end{align}
where
\begin{equation}\label{27}
{\mathbf b}^{\otimes i}:=\underbrace{{\mathbf b}\otimes\cdots
\otimes {\mathbf b}}_i.
\end{equation}
First consider the bulk terms in the time derivative of ${\mathcal
E}^{\mathbf a}_{n}$'s. Demanding that the bulk terms can be
expressed in terms  of the expectations of ${\mathcal E}^{\mathbf
a}_{n}$'s, one arrives at the following condition for the bulk
terms.
\begin{equation}\label{28}
    {\cal A}^{0k0}H=\lambda {\cal A}^{0k0}.
\end{equation}
The condition coming from the left boundary of the block is
\begin{equation}\label{29}
    {\cal A}^{k-1,k-1,0}[{\mathbf 1}^{\otimes k-2}\otimes H +
    +\cdots + H\otimes {\mathbf 1}^{\otimes
    k-2}]= \sum _{{{m,l}\atop{m+l=2k-2}}}C^L_{ml0}{\cal A}^{ml0},
\end{equation}
while the condition coming from the right boundary of the block
\begin{equation}\label{30}
    {\cal A}^{0,k-1,k-1}[{\mathbf 1}^{\otimes k-2}\otimes H +
    \cdots + H\otimes {\mathbf 1}^{\otimes
    k-2}]= \sum _{{{m,l}\atop{m+l=2k-2}}}C^R_{0lm}{\cal A}^{0lm}.
\end{equation}

For the blocks of the length $j$ with $j<k-1$, one needs the
condition
\begin{equation}\label{31}
    {\cal A}^{k-1,j,k-1}[{\mathbf 1}^{\otimes j+k-2}\otimes H +
    \cdots + H\otimes {\mathbf 1}^{\otimes
    j+k-2}]= \sum _{{{m,p,l}\atop{m+p+l=j+2k-2}}}C_{mpl}{\cal A}^{mpl}.
\end{equation}
Finally for the blocks of the length $(L+1-j)$ with $j<k-1$, one
needs the condition
\begin{equation}\label{32}
    {\cal B}^{k-1,j,k-1}[{\mathbf 1}^{\otimes j+k-2}\otimes H +
    \cdots + H\otimes {\mathbf 1}^{\otimes
    j+k-2}]= \sum _{{{m,p,l}\atop{m+p+l=j+2k-2}}}D_{mpl}{\cal B}^{mpl}.
\end{equation}
Now it is easy to write the evolution equation of ${ E}^{\mathbf
a}_n$. For $k-1\leq n\leq L+2-k$,
\begin{align}\label{33}
    \dot { E}^{\mathbf
a}_n =(n-k+1)\lambda { E}^{\mathbf a}_n
    &+\sum_{l=0}^{2k-2}{ E}^{\mathbf a}_{l-k+1+n}C^L_{2k-2-l,l,0}\nonumber\\
    &+\sum_{l=0}^{2k-2}{ E}^{\mathbf a}_{l-k+1+n}C^R_{0,l,2k-2-l}, \qquad k-1\leq n\leq L+2-k.
\end{align}
For $0<n<k-1$,
\begin{equation}\label{33-1}
    \dot { E}^{\mathbf a}_n =\sum_{p=0}^{n+2k-2}
    \sum_{m=0}^{n+2k-2-p}C_{m,p,n+2k-2-p-m}{ E}^{\mathbf a}_{p}, \qquad 0<n<k-1.
\end{equation}
Finally, for $L+2-k<n<L+1$,
\begin{equation}\label{34}
    \dot { E}^{\mathbf a}_n =\sum_{p=0}^{n+2k-2}
    \sum_{m=0}^{n+2k-2-p}D_{m,p,n+2k-2-p-m}{ E}^{\mathbf a}_{L+1-p}, \qquad L+2-k<n<L+1.
\end{equation}
One also has,
\begin{align}\label{36-1}
{ E}^{\mathbf a}_0&=1,\nonumber\\
\dot { E}^{\mathbf a}_{L+1}&=(L+1)\lambda { E}^{\mathbf a}_{L+1}.
\end{align}

Equations (\ref{33}) to (\ref{36-1}) can be solved to obtain a
certain set of correlation functions ($E^{\mathbf a}_n$'s).

\section{A three-site model with annihilation, as an example}
Here, a model with three-site interactions is considered. Each
site may be occupied or be a vacancy. A particle at each site may
be annihilated, and the annihilation rate depends on its
neighboring sites. The interactions are
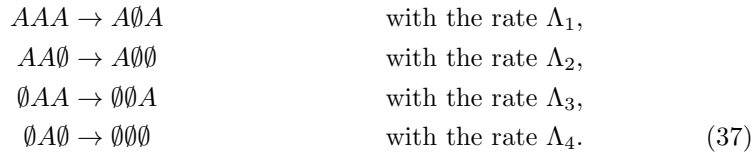
\begin{align}\label{35}
    AAA&\to A\emptyset A& \qquad\hbox{with the rate }\Lambda_1,\cr
    AA\emptyset &\to A\emptyset \emptyset & \qquad\hbox{with the rate }\Lambda_2,\cr
    \emptyset  AA&\to \emptyset \emptyset A &\qquad\hbox{with the rate }\Lambda_3,\cr
    \emptyset A\emptyset & \to \emptyset \emptyset \emptyset &
    \qquad\hbox{with the rate }\Lambda_4.
\end{align}
($A$ and $\emptyset$ denote an occupied site and an empty site,
respectively.) Choosing ${\mathbf a}=(1\quad 0)$, it is seen that
this model fulfills the conditions needed for solvability. This
means that the evolution equations for $E_n$'s with $E_n=\langle
N_1 \cdots N_n\rangle$ are closed. Using the method introduced in
previous sections, one can obtain $E_n$'s. The evolution equation
for $E_n$'s are
\begin{align}\label{37}
 &\dot E_n=-[(n-2)\Lambda_1+\Lambda_2+\Lambda_3)]E_n+(\Lambda_2+\Lambda_3-2\Lambda_1)E_{n+1},\quad
 2\leq n\leq L-1\nonumber\\
&\dot
E_1=-\Lambda_4E_1-(\Lambda_2+\Lambda_3-2\Lambda_4)E_2+(\Lambda_2+\Lambda_3-\Lambda_1-\Lambda_4)E_3,\nonumber\\
&\dot
E_L=-(\Lambda_2+\Lambda_3)E_L+(\Lambda_2+\Lambda_3-2\Lambda_1)E_{L+1}\nonumber\\
&\dot E_{L+1}=-(L+1)\Lambda_1E_L
\end{align}
To simplify the model, assume $\Lambda_1=0$. This means that a
block of particles, can only be altered at the boundaries.
Defining
\begin{align}\label{38-0}
    \alpha:&=\Lambda_2+\Lambda_3,\cr
    \beta:&=\Lambda_4,
\end{align}
(\ref{37}) is rewritten as
\begin{align}\label{38}
 &\dot E_n=\alpha(E_{n+1}-E_n),\quad
 2\leq n\leq L\cr
&\dot E_1=\beta(E_2-E_1)+(\alpha-\beta)(E_3-E_2),\cr  &\dot
E_{L+1}=0.
\end{align}
There are two cases.

\noindent \textbf{i}) $\alpha=0$. In this case, all the $E_n$'s
for $n\ne 1$ are constant and
\begin{equation}\label{39}
    E_1(t)=e^{-\beta t}[E_1(0)-2E_2(0)+E_3(0)]+2E_2(0)-E_3(0).
\end{equation}

\noindent \textbf{ii}) $\alpha\ne 0$. In this case, $E_{L+1}$ is
constant. Using this, one can obtain $E_L$, and then $E_{L-1}$,
..., and $E_2$. The result is

\begin{equation}\label{42}
E_{L-k}=E_{L+1}+\left[\sum_{j=0}^k\frac{\gamma_{L-k+j}}{j!}\,
(\alpha\,t)^j\right] e^{-\alpha\, t},
\end{equation}
where $\gamma_k$'s are arbitrary constants to be determined from
the initial conditions. $E_1$ is also easily determined from the
second equation of (\ref{38}).\\
\\
{\bf Acknowledgement} The authors would like to thank
M.~Kaboutari, for useful discussions.


\newpage
\begin{thebibliography}{99}
\bibitem{ScR}  G. M. Sch\"{u}tz; ``Exactly solvable models for many-body
               systems far from equilibrium'' in ``Phase
               transitions and critical phenomena, vol. \textbf{19}'',
               C. Domb \& J. Lebowitz (eds.), (Academic
               Press, London, 2000).
\bibitem{ADHR} F. C. Alcaraz, M. Droz, M. Henkel, \& V. Rittenberg;
               Ann. Phys. \textbf{230} (1994) 250.
\bibitem{KPWH} K. Krebs, M. P. Pfannmuller, B. Wehefritz, \&
               H. Hinrichsen; J. Stat. Phys. \textbf{78}[FS] (1995) 1429.
\bibitem{HS1}  H. Simon; J. Phys. \textbf{A28} (1995) 6585.
\bibitem{PCG}  V. Privman, A. M. R. Cadilhe, \& M. L. Glasser; J. Stat.
               Phys. \textbf{81} (1995) 881.
\bibitem{HOS1} M. Henkel, E. Orlandini, \& G. M. Sch\"utz; J. Phys.
               \textbf{A28} (1995) 6335.
\bibitem{HOS2} M. Henkel, E. Orlandini, \& J. Santos; Ann. of Phys.
               \textbf{259} (1997) 163.
\bibitem{AL}   A. A. Lushnikov; Sov. Phys. JETP \textbf{64} (1986) 811
               [Zh. Eksp. Teor. Fiz. \textbf{91} (1986) 1376].
\bibitem{AKK}  M. Alimohammadi, V. Karimipour, \& M. Khorrami; Phys. Rev.
               \textbf{E57} (1998) 6370.
\bibitem{RK}   F. Roshani \& M. Khorrami; Phys. Rev. \textbf{E60} (1999) 3393.
\bibitem{RK2}  F. Roshani \& M. Khorrami; J. Math. Phys. \textbf{43} (2002) 2627.
\bibitem{AKK2} M. Alimohammadi, V. Karimipour, \& M. Khorrami; J. Stat.
               Phys. \textbf{97} (1999) 373.
\bibitem{AAMS} A. Aghamohammadi, A. H. Fatollahi, M. Khorrami, \& A.
               Shariati; Phys. Rev. \textbf{E62} (2000) 4642.
\bibitem{AM1}  A. Aghamohammadi \& M. Khorrami; J. Phys. \textbf{A33} (2000) 7843.
\bibitem{AA}   M. Alimohammadi, \& N. Ahmadi; Phys. Rev. \textbf{62} (2000) 1674.
\bibitem{RK3}  F. Roshani \& M. Khorrami; Phys. Rev. \textbf{E64} (2001) 011101.
\bibitem{RK4}  F. Roshani \& M. Khorrami; Eur. Phys. J. \textbf{B36} (2003) 99.
\bibitem{BDb}  M. A. Burschka, C. R. Doering, \& D. ben-Avraham;  Phys.
               Rev. Lett. \textbf{63} (1989) 700.
\bibitem{BDb1} D. ben-Avraham;  Mod. Phys. Lett. \textbf{B9} (1995)
               895.
\bibitem{BDb2} D. ben-Avraham; in ``Nonequilibrium Statistical
               Mechanics in One Dimension'', V. Privman (ed.), pp 29-50
               (Cambridge University press,1997).
\bibitem{BDb3} D. ben-Avraham; Phys. Rev. Lett. \textbf{81} (1998)
               4756.
\bibitem{Mb}   T. Masser, D. ben-Avraham; Phys. Lett. \textbf{A275} (2000) 382.
\bibitem{AKA}  M. Alimohammadi, M. Khorrami, \& A. Aghamohammadi;
               Phys. Rev. \textbf{E64} (2001) 056116.
\bibitem{HH}   M. Henkel \& H. Hinrichsen; J. Phys. \textbf{A34}, 1561
               (2001).
\bibitem{KAA}  M. Khorrami, A. Aghamohammadi, \& M. Alimohammadi;
               J. Phys. \textbf{A36} (2003) 345.
\bibitem{MB}   M. Mobilia \& P. A. Bares; Phys. Rev. \textbf{E64} (2001) 066123.
\bibitem{AAK}  A. Aghamohammadi, M. Alimohammadi, \& M. Khorrami;
               Eur. Phys. J. \textbf{B31} (2003) 371.
\end{thebibliography}
\end{document}